\documentclass [titlepage, nofootinbib]{revtex4}
\usepackage{amssymb}
\usepackage{graphicx}
\usepackage{amsmath}
\bibpunct[, ]{[}{]}{,}{a}{,}{,}
\newcommand{\Pheno}{{m}}
\newcommand{\ab}{{\mathcal A}}  

\newcommand{\qq}{{p}}
\newcommand{\pp}{{q}}
\newcommand{\qE}{{\hat{\qq}}}
\newcommand{\pE}{{\hat{\pp}}}
\newcommand{\PP}{\mathcal{Q}}
\newcommand{\PE}{\mathcal{\hat Q}}

\newcommand{\MutRate}{{\mu_0}} 
\newcommand{\Mut}{{U}} 
\newcommand{\threshold}{{\Pheno_0}}

\begin{document}

\title{Analytical study of the effect of recombination on evolution via DNA shuffling}
\author{Weiqun Peng$^*$}
\author{Herbert Levine$^*$}
\author{Terence Hwa$^*$}
\author{David A. Kessler$^{\dagger}$}
\affiliation{$*$Center for Theoretical Biological Physics, University of
California San Diego, 9500 Gilman Drive, La Jolla CA 92093, USA}
\affiliation{${\dagger}$Department of Physics, Bar-Ilan University, Ramat-Gan,
Israel}
\begin{abstract}
We investigate a multi-locus evolutionary model which is based on
the DNA shuffling protocol widely applied in \textit{in vitro}
directed evolution. This model incorporates selection,
recombination and point mutations. The simplicity of the model
allows us to obtain a full analytical treatment of both its
dynamical and equilibrium properties, for the case of an infinite
population. We also briefly discuss finite population size
corrections.
\end{abstract}
\maketitle
Recombination of genetic information is a widespread phenomena in
both prokaryotes and eukaryotes.  In prokaryotes, recombination
occurs in the forms of transformation, virus aided transduction,
and conjugation. In eukaryotes, recombination between homologous
sequences is a fundamental component underlying sexual
reproduction~\nocite{REF:Lodish2000}({\it e.g.},
L{\scriptsize{ODISH}} {\it et.al.} $1999$). An enormous body of
research has been devoted to understanding the evolutionary
benefit of recombination in various circumstances. This effort has
led to some general understanding of the circumstances under which
recombination helps facilitate evolution, although many important
questions still remain open~\nocite{OttoB02}(O{\scriptsize{TTO}}
and L{\scriptsize{ENORMAND}} $2002$).

Inspired by recombination in natural evolution, and propelled by
advances in biotechnology, recombination has been employed in
\textit{in vitro} molecular evolution experiments to develop novel
proteins and DNA
sequences~\nocite{KurtzmanGVJHP01,FarinasBA01}({\it e.g.},
K{\scriptsize{URTZMAN}} {\it et.al.} $2001$;
F{\scriptsize{ARINAS}} {\it et.al.} $2001$). This family of
evolutionary protocols, called DNA shuffling (or molecular
breeding)~\nocite{REF:Stemmer1994,
REF:Stemmer19942}(S{\scriptsize{TEMMER}} $1994$a; $1994$b), has
been shown experimentally to produce, in terms of the rate of
evolutionary progression and final product quality, far superior
results as compared to conventional directed evolution methods
using only mutagenesis~\nocite{KurtzmanGVJHP01,FarinasBA01}({\it
e.g.}, K{\scriptsize{URTZMAN}} {\it et.al.} $2001$;
F{\scriptsize{ARINAS}} {\it et.al.} $2001$). In addition to its
widespread practical applications, DNA shuffling (as an approach
of {\it in vitro} evolution) has been used to mimic natural
evolutionary processes and predict possible evolutionary
pathways~\nocite{OrenciaYNSS01,BarlowH021}(O{\scriptsize{RENCIA}}
{\it et.al.} $2001$; B{\scriptsize{ARLOW}} and H{\scriptsize{ALL}}
$2002$).

In spite of its enormous significance, theoretical understanding
of DNA shuffling has been lacking. In this paper, we investigate
DNA shuffling from the perspective of evolutionary modelling.
Specifically, we aim to find out, quantitatively and analytically,
the benefit that the extra step of recombination provides in the
evolutionary process, as well as such aspects as the role of
mutation and finite population effects.

Compared to other evolutionary models, DNA shuffling has two
unique aspects that render it attractive to study. First, the methods
of biotechnology enable a unique recombination scheme that goes well
beyond the classical one, i.e., two parents with at most a few
crossovers. This multi-parent multi-crossover nature of DNA
shuffling, as we shall see, makes it more effective, and also,
incidentally, dramatically simplifies the analytical treatment.
Second, in the setting of DNA shuffling the relationship between
genotype, phenotype and fitness is relatively clear and well-defined.
In addition, such parameters as selection strength, mutation rate and
the amount of recombination are all experimentally controllable.
These two characteristics should allow theoretical results to be
tested directly against experiments.

Specifically, we propose a simple model that incorporates three
basic ingredients: selection, recombination and point mutations.
In order to facilitate comparison with various existing
evolutionary models, we present our ideas in the language of
population genetics. In this language, we study a haploid
multi-locus model with multi-parent free recombination, and
subject to dynamical truncation selection, wherein the fitness of
a genotype depends on the population state. We obtain analytical
results for both the dynamics and the equilibrium properties, and
gain insights into not only why, but also \textit{how} exactly
recombination works. This is one of the rare cases in population
genetics where exact results can be obtained for a multi-locus
model~\nocite{KesslerLRT97}(K{\scriptsize{ESSLER}} {\it et.al.}
$1997$).

\begin{center}
\section*{THE MODEL}
\end{center}
DNA shuffling involves a directed evolution process wherein a
library of homologous DNA sequences are subject to rounds of
competitive selection and \textit{in vitro} recombination with
multiple parents and multiple crossovers~\nocite{REF:Stemmer1994,
REF:Stemmer19942}(S{\scriptsize{TEMMER}} $1994$a; $1994$b). DNA
shuffling is a discrete process with non-overlapping generations,
each round of which involves selection, recombination and point
mutation, and is finished by amplification via the polymerase
chain reaction (PCR) of the population back to it original size.
The typical selection scheme in DNA shuffling experiments is
truncation selection, also known as breeding selection, where only
a fixed portion of the population (e.g., the top ten percent) is
chosen to be retained to participate in later rounds.  In
truncation selection, whether a particular member of the
population is selected or not depends on whether the desired trait
exceeds a certain threshold set by the population as a whole and
by the selection strength (i.e, the fraction of selected
population).  As opposed to other evolutionary scenarios, there is
no advantage to being better than the cutoff threshold -- there is
no pressure to excel. Mathematically, this has a dramatic effect
on the dynamics, in that the effect of population size is much
weaker, as the rare ``superstars'' found only in a large
population do not skew the results.

The recombination step typically involves random fragmentation of
homologous DNA sequences by DNase digestion and repeated cycles of
re-assembly via a self-primed polymerase chain
reaction~\nocite{REF:Stemmer1994,
REF:Stemmer19942}(S{\scriptsize{TEMMER}} $1994$a; $1994$b).
Recombination produces chimeras with a controllable average
fragment size (of order $10$ base pairs or above). Point mutation
is incorporated via the recombination and amplification steps
where PCR is utilized and is naturally error-prone. To date, the
rate of point mutation has been kept very low and only single
nucleotide substitution is assumed to be involved. Thus, sequence
diversity has come mostly from the diversity present in the
initial library instead of being generated by point mutation. This
lessens the deleterious effects associated with high rates of
mutagenesis.
but  ultimately
limits the usefulness of the method. A proper balance of
recombination and mutation would lead to more optimal results.

Despite its enormous practical success, theoretical analysis of
the evolutionary dynamics of DNA shuffling has thus far been
lacking. Sun~\nocite{REF:Sun1999}($1999$), Moore and
Maranas~\nocite{REF:Moore2000}($2000$) proposed predictive models
for various experimental steps of DNA shuffling experiments,
addressing issues of recombination efficiency and distribution of
fragment size during the reassembly involved in a single round of
evolution. We concern ourselves instead with the evolutionary
consequences of multiple rounds.

To make our discussion concrete, we envision the competitive
evolution of a library of DNA sequences, selected via binding
affinity to certain proteins; an example of such a system is the
DNA-histone
interaction~\nocite{ThastromLWCKW99}(T{\scriptsize{HASTROM}} {\it
et.al.} $1999$). As discussed elsewhere~\nocite{VonhippelB86,
Peng02, GerlandH02}(V{\scriptsize{ON}} H{\scriptsize{IPPEL}} and
B{\scriptsize{ERG}} $1986$; P{\scriptsize{ENG}} {\it et.al.}
$2003$), selection achieved via thermodynamic binding can be
mimicked to a high level of accuracy by the simple truncation
selection approach.
\setlength{\unitlength}{1mm}%
\begin{figure}
\includegraphics*[  height=2.0in]{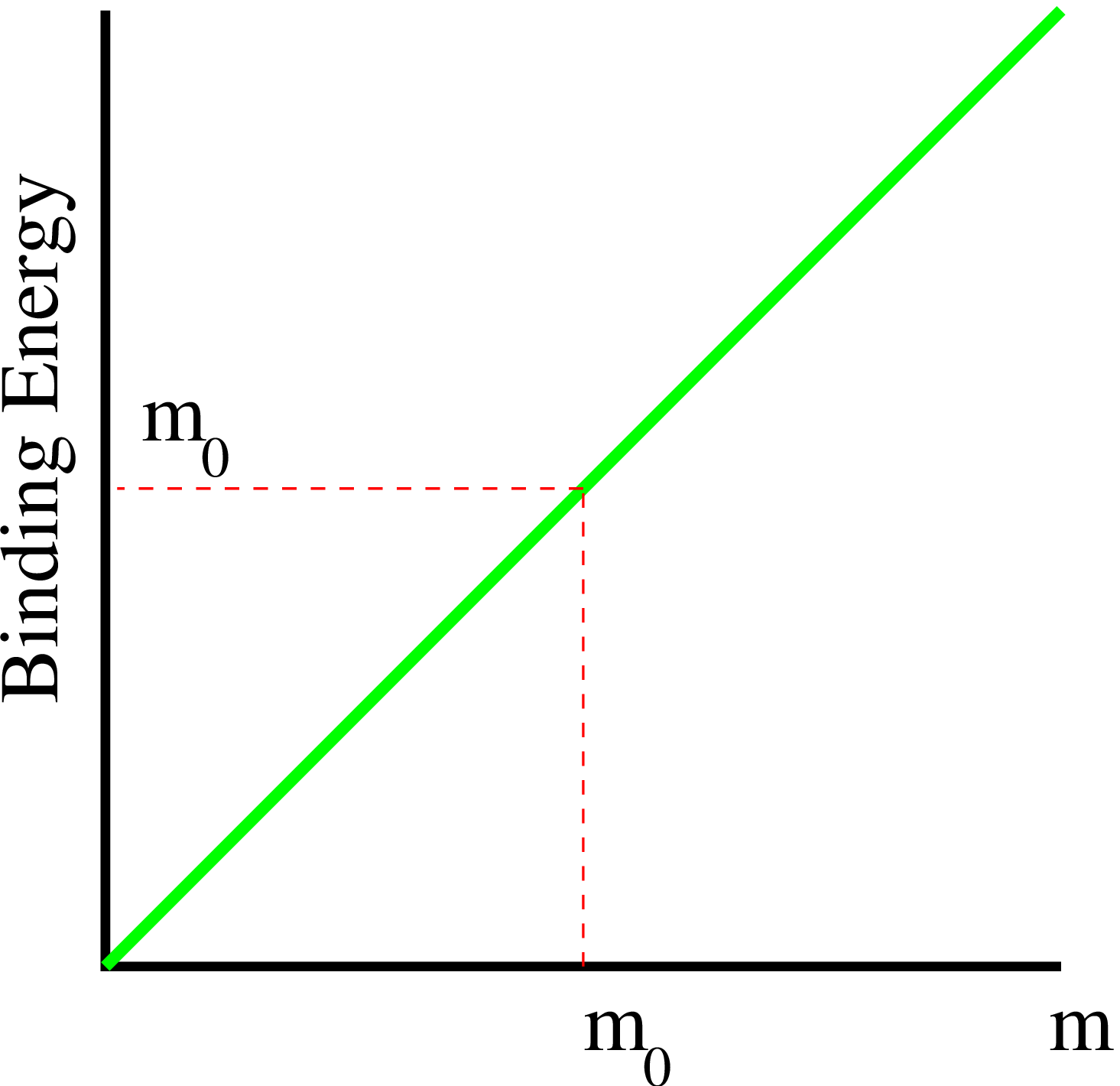}
\includegraphics*[  height=2.1in]{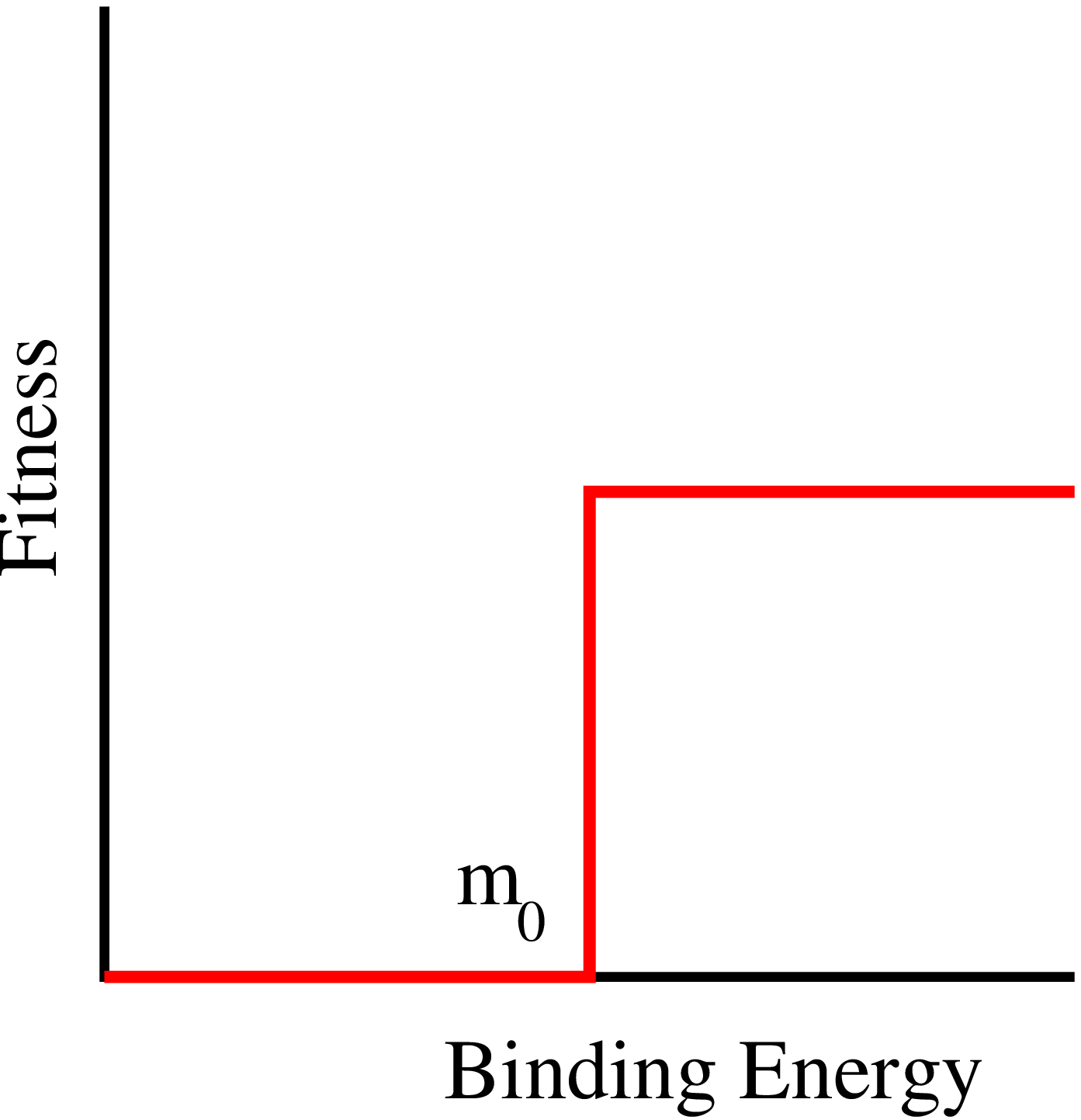}
\caption{\protect\flushing (a) Phenotypic landscape: Binding energy
is exactly the number of favorable active sites (denoted by $m$).
Dash line indicates the corresponding selection threshold $m_{0}$ at a particular
stage of evolution. (b) The fitness has a dynamic truncation landscape.
Sequences with binding energy below the threshold $m_{0}$  are
discarded, whereas those with binding energy above the threshold are
retained to reproduce with equal rate.
\label{FIG:PhenotipicLandcape}}
\end{figure}
In order to facilitate a theoretical treatment, we construct
simplified models of the recombination and selection steps. Let us
describe the selection step first. For selection, we need a model
that connects genotype and phenotype, which is in this case the
binding energy of the DNA to the protein, as well as a model that
specifies selection on the phenotype. We adopt the simplest
relationship between genotype and phenotype. We assume that each
nucleotide contributes to the binding energy independently and
additively.  Each nucleotide can be one of the $\ab$($=4$)
nucleotides (representing A, C, G and T), among which one is
favorable and the rest are equally deleterious. For simplicity, we
denote the contribution of a specific site to the binding energy
$1$ ($0$) if therein sits a favorable (deleterious) nucleotide.
We will refer to a favorable (deleterious) nucleotide as a match
(mismatch) with respect to the optimal sequence. This formulation
can be derived from 
a two-state model for protein-DNA
binding~\nocite{VonhippelB86,GerlandH02}(V{\scriptsize{ON}}
H{\scriptsize{IPPEL}} and B{\scriptsize{ERG}} $1986$;
G{\scriptsize{ERLAND}} and H{\scriptsize{WA}} $2002$). The binding
energy of the sequence is simply the number of sites with
favorable nucleotides.

Fig.~(\ref{FIG:PhenotipicLandcape}a) shows schematically the shape
of our phenotypic landscape. Of course, our model reflects just
this simplest possibility. In reality, the phenotypic landscape
could be much more complicated and is rarely known; in fact, a
significant advantage of directed evolution over rational design
is that this kind of knowledge is not
necessary~\nocite{Arnold01}({\it e.g.} A{\scriptsize{RNOLD}}
$2001$). Our strategy here is to study the simplest non-trivial
model available, obtain a thorough understanding, with the hope of
proceeding to more complicated situations to see which results
obtained in the simple model are general and which are model
specific. Having specified the phenotypic landscape, we next
describe the selection protocol. Selection acts on the binding
energy of the sequences.  As noted above, in (molecular) breeding,
truncation selection is used. We define the selection strength via
the fraction $\phi $ of selected members of the total population.
Suppose we have a distribution of population $P_{\Pheno}$ in terms
of the binding energy $\Pheno(\Pheno=0,1,\ldots ,L)$, where $L$ is
the total number of active sites. The threshold $\threshold $ is
self-consistently determined by
\begin{equation}
\phi =\alpha P_{\threshold }+\sum _{\Pheno=\threshold+1}^{L}P_{\Pheno},
\end{equation}
where $\alpha $ in the first term takes into account of the
partial selection on the threshold state $\threshold $. $\phi $
varies from $\phi \lesssim 1$ (relatively weak selection) to small
$\phi$ (relatively strong selection). For those sequences whose
number of matches are above (below) $\threshold $, their fitness
is $1$ ($0$), and they are selected (discarded) {[}see
Fig.~(\ref{FIG:PhenotipicLandcape}b){]}.  In the truncation
selection scheme every member's fitness is collectively and
dynamically determined by both the phenotypic distribution of the
population $P_{\Pheno}$ and selection strength $\phi $. In
contrast, most fitness landscapes studied prescribe for each
genotype a preset fitness value and its effective fitness value is
simply its own fitness over the mean
fitness~\nocite{REF:Crow1970}(C{\scriptsize{ROW}} and
K{\scriptsize{IMURA}} $1970$). Truncation selection generates
correlations (i.e., linkage) between loci, hence it is
epistatic~\nocite{OttoB01, ShnolK93}(O{\scriptsize{TTO}} and
B{\scriptsize{ARTON}} $2001$; S{\scriptsize{HNOL}} and
K{\scriptsize{ONDRASHOV}} $1993$).

We note in passing that the term truncation selection has also been
used to mean a {\it fixed} step-like landscape. In population
genetics literatures, a fixed step-like landscape is also called {\it
hard} truncation selection, and the selection scheme we employed is
sometimes termed {\it soft} truncation selection.  Due to the
dynamical nature of the selection scheme, using fitness as a
yardstick for the evolution is not very useful;  for example, the
mean fitness of the population is always $\phi$ by definition. As a
consequence, we will focus on the evolution of the phenotypic
distribution, i.e., the distribution of binding energies,
instead of the fitness distribution.

We now turn to the discussion of recombination. Our basic approach
here will be to make a substantial simplification of the actual
situation and assume perfect recombination between all nucleotides.
Namely, in building each new sequence after selection, each
nucleotide independently samples the nucleotides at the corresponding
sites in the entire selected population.  Formally, this amounts to
assuming that the fragmentation and reassembly process is repeated
often enough such that there is no linkage between any of the
nucleotides. This is undoubtedly false in detail for the experiments
done to date, but we shall see that this approximation is quite good
for describing the result of a scenario involving the more feasible
case of a finite number of crossovers.  The advantage of this
``maximal'' recombination is that it allows for an analytic solution
of the model, as will be presented in the remainder of this paper.
Finally, we assume simple point mutation for each site. Namely, each
nucleotide is independently subject to mutation rate $\MutRate$ per
generation. This process includes both beneficial and deleterious
mutations.

\setlength{\unitlength}{1mm}%
\begin{figure}
\includegraphics*[ height=2.5in]{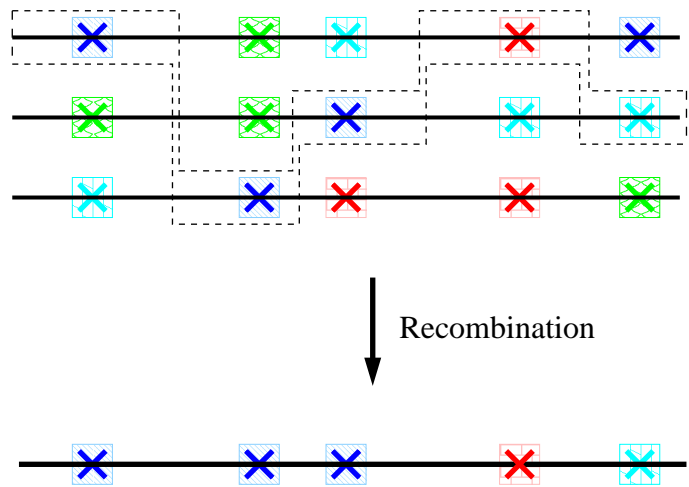}
\caption{\protect\flushing Maximal recombination under a general
interpretation.  Each line represents a DNA sequence. Crosses
represents active sites with variable length fixed regions in
between. Crosses with different background patterns represent
different phenotypes for active sites. In building a new sequence,
each active site independently samples the corresponding sites in the
entire selected population. \label{FIG:recombinationscheme}}
\end{figure}

It is worth pointing out that our approach to recombination may
have more general applicability than merely as an approximation
for the DNA binding problem. In many cases of DNA shuffling, the
initial library of genes coding for an interesting protein is
chosen from closely related species so as to ensure proven
functionality and sufficient homology to allow recombination. Let
us imagine that the majority of the nucleotides along those DNA
sequences are already optimal, and (non-synonymous) mutations of
these nucleotides are lethal. Hence, we can consider these
nucleotides to be fixed during the entire evolutionary process and
ignore them except insofar as they provide enough homology so that
overlapping single-stranded fragments from different DNA sequences
can anneal with each other (and do not anneal with other fragments
by accident) during the self-primed PCR re-assembly step
(S{\scriptsize{TEMMER}} $1994$a;
$1994$b)~\nocite{REF:Stemmer19942,REF:Stemmer1994}. For the
remaining {\it active} sites, we assume that they are (a)~ far
enough from each other so that recombination happens freely
between any two; (b)~they are subject to point mutations with rate
$\MutRate $ per nucleotide per generation. Therefore, each active
site, along with its fixed flanking homologous regions (whose size
or exact delineation do not matter), constitutes a segment. To
build a new sequence from recombination, a number of overlapping
fragments (which when put together cover the entire sequence) are
assembled, with each fragment obtained from randomly sampling the
corresponding fragments (i.e., having the same active site) in the
selected population. This idea is depicted in
Fig.~(\ref{FIG:recombinationscheme}) and is an exact realization
of our maximal recombination model. Of course, one has to be
careful to intersperse an experimental selection step that will
eliminate all lethal variants (due to mutation) before proceeding
to an actual selection based on useful variation. To proceed, we
would then have to explicitly take into account the transformation
from gene sequence to amino acid, as the selection would be on the
basis of some desired activity of the protein. We do not pursue
this line of investigation any further in this work.

Finally, we can rephrase our model in terms of the standard language
of population genetics. Each site can be referred to as a locus, which
can have $\ab $ alleles, one favorable (match) and the rest
(mismatches) equally unfavorable. The set of all $L$ loci form a
chromosome. The recombination scheme is such that the allele of each
locus of every new chromosome is chosen by randomly sampling the
alleles at the corresponding locus of all the selected
chromosomes. The fitness value of each chromosome is either $1$ ($0$),
when the number of matches it has is above (below) the threshold
$m_{0}$. For clarity, we list the correspondence in
Table~\ref{JargonCorrespondence}.

\vspace{.5cm}
\begin{table}
\begin{tabular}{|c|c|}
\hline DNA-protein binding& Population Genetics\\ \hline \hline DNA
sequence& chromosome\\ \hline nucleotide& locus\\ \hline binding
energy& phenotypic value:matches\\ \hline selection via binding to protein& dynamical truncation selection\\\hline
\end{tabular}
\caption{ The correspondence between two languages: directed molecular evolution
and population genetics.
\label{JargonCorrespondence}}
\end{table}

\begin{figure}
\includegraphics*[  height=2.5in]{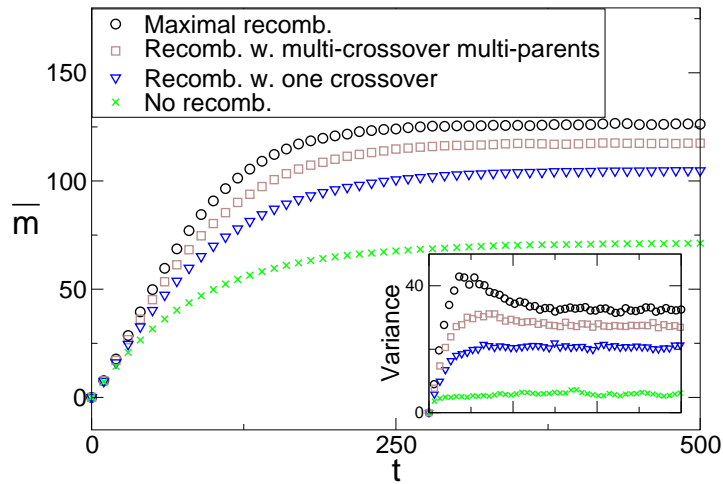}
\caption{\protect \flushing Comparison of evolutionary trajectories
for four different evolutionary protocols under weak selection. Shown
are simulation results of the evolutionary trajectory of the average
number of matches ${\overline m}$ for the population, with an initial
clonal population with no matches at any chromosomes. x: no
recombination; Triangle: recombination with one random crossover;
Square: recombination with multiple crossover and multiple parents,
using a per bond crossover probability of $0.025$. Circle: maximal
recombination.  Shown in the inset are the corresponding variances of
the distributions versus time, ordered in the same way as the matches
vs. generations curves. Here sequence length $L=170$, selection
strength $\phi=.9$, mutation rate $\MutRate=.01$ and population size
$N=10^4$. \label{FIG:EnergyEvo}}
\end{figure}

In genetic algorithms and evolutionary strategies, a research area
in computer science where the principles of evolution are employed
to find optimal solutions to complex
problems~\nocite{REF:Beyer2001}(B{\scriptsize{EYER}} $2001$), a
similar setting has been investigated by M{\"u}hlenbein and
Schlierkamp-Voosen~\nocite{REF:Muhlenbein1993}($1993$), mostly via
computer simulation. Various aspects of classical breeding have
also been studied in population
genetics~\nocite{REF:Crow1970}(C{\scriptsize{ROW}} and
K{\scriptsize{IMURA}} $1970$). Kimura and Crow investigated the
response of individual loci to selection in one round, under
linkage-free
condition~\nocite{REF:Crow1970,REF:Kimura1978}(C{\scriptsize{ROW}}
and K{\scriptsize{IMURA}} $1970$; K{\scriptsize{IMURA}} and
C{\scriptsize{ROW}} $1978$).
Kondrashov~\nocite{REF:Kondrashov1995}($1995$) studied a model
with a different type of dynamical fitness landscape,  where the
fitness value of a genotype depends only on the difference between
its phenotypic value and the mean of the phenotypic distribution,
in units of variance of the phenotypic distribution. Along with
the assumption of deleterious mutations, conventional
recombination, and a Gaussian distribution before selection
(which, as we shall see, is in fact equivalent to using maximal
recombination), he derived evolutionary recursion relations and
obtained analytical expressions that characterize the equilibrium
position.  We shall make connections with these works in
appropriate places.\begin{center}\end{center}
\begin{center}
\section*{RESULTS} 
\end{center}
{\bf Dynamics of maximal recombination}\\
To summarize the above discussion, in the language of population
genetics which we will adopt from here on, we have a haploid
population of $N$ chromosomes each with $L$ loci, and we study the
evolution with dynamical truncation selection, maximal
recombination and point mutation as specified above. We fix the
order of operation to be selection, recombination and mutation. At
each step the population size $N$ is kept constant. This model is
easily simulated on a computer. Fig.~(\ref{FIG:EnergyEvo}) shows
simulation results for evolutionary trajectories of the average
and variance of the match distribution under weak selection. For
comparison, we show results from four different evolutionary
protocols: no recombination; single random crossover; multiple
crossovers and multiple parents (which is the situation closest to
experiments); and maximal recombination. It is evident that
recombination improves both the dynamics and equilibrium state as
compared to the case with no recombination. For our phenotypic
landscape and selection scheme, the more recombination the better.
Furthermore, the maximal recombination scheme captures the
essential effect of recombination and provides a reasonable
approximation to the more readily achievable case of multiple
crossovers from multiple parents. In the alternative case of
strong selection (not shown), all evolutionary protocols propel
the population to an equilibrium state very close to the optimal
state; however the population reaches the equilibrium state much
faster when recombination is applied.

\setlength{\unitlength}{1mm}

In our analytical work we will assume, unless specifically noted
otherwise, the population size $N$ to be very large so that random
genetic drift is negligible; we will briefly discuss finite size
effects at the end. We focus on the evolution of such macroscopic
characteristics of the population as mean and variance, as opposed to
``microscopic'' properties such as the fate of individual mutations.
As already noted, the fitness distribution itself does not describe
the evolution. We instead study the evolution of the phenotypic (match)
distribution, focusing on its first and second moments. For
simplicity we assume linkage equilibrium at the beginning of
evolution. This is in fact not a stringent assumption, since linkage
equilibrium is achieved anyway right after one round of
recombination.

Even before going into any details of the analysis, it is not
difficult to understand why recombination is highly beneficial in
this model. Selection, as it operates on the phenotype,
which in this case is the total number of matches,
introduces linkage. After the selection step, recombination
immediately restores total linkage equilibrium. resulting into a
broader distribution [see Fig.~(\ref{FIG:EnergyEvo}) inset]. A
broader distribution means better response to subsequent selection
and hence faster evolution. The following analysis describes how this
works in a fully quantitative way, in the maximal recombination
model.

{\bf Dynamics without mutation:}
We start from the simplest case where the mutation rate $\MutRate$
is set to be $0$.  We assume that each locus has the same binary
distribution characterized by the probability of being favorable
(i.e., a match) $\qq(0)$ $(\neq 0)$. This is in fact the case most
relevant to the DNA shuffling experiments to date, where
$\MutRate$ is kept extremely small and the diversity is almost
entirely provided by the diversity which existed in the initial
library~\nocite{KurtzmanGVJHP01}(K{\scriptsize{URTZMAN}} {\it
et.al.} $2001$). Starting from the homogeneous initial condition,
we expect that every locus follows the same evolution trajectory,
as the evolution dynamics preserve permutation symmetry of
different loci. With this in mind, we focus on the evolution of
the probability $\qq(t)$ for one locus to be favorable at the end
of generation $t$.  The phenotypic probability distribution for
the number of matches of a chromosome at the end of round $t$ is
then a binomial distribution characterized by mean $L\qq(t)$.
Assuming that $L$ is large (For a more accurate treatment without
the assumption of large $L$, see
Appendix~\ref{APP:ExactDynamics}), the binomial distribution is
well approximated by a Gaussian distribution $P(m,t)$
:
\begin{eqnarray}
P(m,t)&=&\frac{1}{\sqrt{2\pi \sigma ^{2}(t)}}\exp
\left(-\frac{(m-{\overline{\Pheno}}(t))^{2}}{2\sigma ^{2}(t)}\right)
\label{EQ:Gaussian}\\
{\overline{\Pheno}}(t)&=&L\qq(t)
\label{Eq:GaussianMean}\\
\sigma ^{2}(t)&=&L\qq(t)\,(1-\qq(t))
\label{Eq:GaussianVariance}
\end{eqnarray}

\begin{figure}
\includegraphics*[  height=2in]{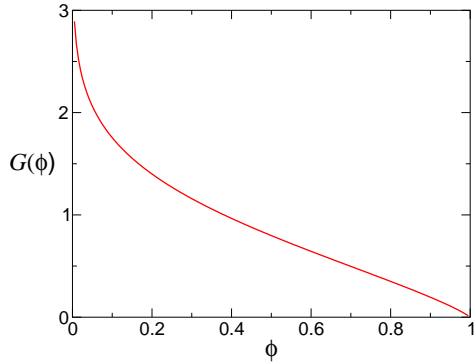}
\caption{\protect
\flushing \label{FIG:Gbar} The selection factor $G(\phi)$. $G(\phi)$
diverges at strong selection $\phi \to 0$ and goes to zero at weak
selection $\phi \to 1$.}
\end{figure}

Given such a distribution at the end of generation $t$, we now discuss
step by step the effects on the distribution due to various operations
in the evolutionary protocol. In generation $t+1$, selection cuts out
the low $m$ tail of the Gaussian distribution.
It is straightforward to calculate that
the mean match number of the selected population
${\overline{\Pheno}}_{\textrm{s}}$ is
\begin{equation}
{\overline{\Pheno}}\rightarrow
{\overline{\Pheno}}_{\textrm{s}}={\overline{\Pheno}}+\sigma G(\phi ).
\label{EQ:Selection}
\end{equation}
The new mean can thus be expressed as the old mean plus an improvement due
to selection, which is simply the product of the old standard
deviation and a factor $G(\phi)$ that solely encodes the strength of
selection. Here $G(\phi)$ is the mean for the normalized distribution
resulting from a standard Gaussian distribution truncated by a
$1-\phi$ fraction taken off the tail, namely
\begin{equation}
G(\phi )
=\frac{1}{\sqrt{2\pi }\phi }\exp\left(
-\frac{1}{2}X(\phi)^{2}\right),
\label{EQ:GQ}
\end{equation}
 where $X(\phi)$ is the match threshold
defined through
\begin{equation}
\phi =\int _{X(\phi)}^{\infty }\frac{dx}{\sqrt{2\pi
}}\exp\left(-\frac{1}{2}x^{2}\right).
\label{EQ:X}
\end{equation}
Fig.~(\ref{FIG:Gbar}) shows the behavior of $G(\phi)$. For
$0.3<\phi<1$, $G(\phi)$ is approximately a linear function of $\phi$
(with slope roughly $-1.5$), and $G(\phi )\rightarrow (1-\phi
)\sqrt{2\ln\, (1-\phi)^{-1}}$ as $\phi \to 1$. In the strong
selection limit, i.e., $\phi \rightarrow 0$, $G(\phi )\rightarrow
\sqrt{2\ln \phi^{-1} }$, which diverges.
\setlength{\unitlength}{1mm}%

After selection, the recombination step restores the independence of
each locus. The population distribution of matches returns to a
binomial (Gaussian) distribution characterized by mean
${\overline{\Pheno}}_{\textrm{s}}$.
\begin{eqnarray}
{\overline{\Pheno}}_{\textrm{r}} & = & {\overline{\Pheno}}_{\textrm{s}}
\label{EQ:SelectionR}\\
\sigma _{\textrm{r}}^{2} & = &
{\overline{\Pheno}}_{\textrm{r}}(1-{\overline{\Pheno}}_{\textrm{r}}/L)
\end{eqnarray}
Because of the independence of each locus, we can reexpress
Eq.~(\ref{EQ:SelectionR}) in terms of individual match probability
$\qq_{\textrm r}$ and $\qq(t)$:
\begin{equation}
\qq_{\textrm{r}}=\qq(t)+\frac{\sqrt{\qq(t)(1-\qq(t))}}{\sqrt{L}}G(\phi ),
\label{EQ:qrecomb}
\end{equation}
and the variance of the distribution is simply
$L\qq_{\textrm{r}}(1-\qq_{\textrm{r}})$.  Eq.~(\ref{EQ:qrecomb})
has two features: First, the scaled combination of ${G(\phi
)}/{\sqrt{L}}$ is the single control parameter. Second, the change
of $\qq$ is proportional to the square root of the variance,
instead of the variance itself as in the case of a fixed
landscape~\nocite{REF:Crow1970}(C{\scriptsize{ROW}} and
K{\scriptsize{IMURA}} $1970$). This results in a slower evolution
for populations that are in the middle of the landscape, but
results in a speed-up near the $\qq=0$ and $\qq=1$ absorbing
states.

In the case of weak selection, Eq.~(\ref{EQ:qrecomb}) can be approximated by
its continuous-time version:
\begin{equation}
\frac{d\qq}{dt}=\frac{G(\phi )}{{\sqrt{L}}}\sqrt{\qq(1-\qq)}.
\label{EQ:SimShuffling}
\end{equation}
The evolutionary dynamics is governed by two fixed points: $\qq=0$ and
$\qq=1$. $\qq=0$ is a trivial unstable fixed point. When $\qq(0)>0$,  the
population moves toward $\qq=1$ with following time dependence,
\begin{equation}
\qq(t)=\frac{1}{2}\left[1-\sin (\beta -\frac{G(\phi )}{{\sqrt{L}}}t)\right],
\label{EQ:SimShufflingSolution}
\end{equation}
 where $\beta =\sin ^{-1}(1-2\qq_{0})$. Eq.~(\ref{EQ:SimShuffling}) and
its solution have also been derived in M\"{u}hlenbein and
Schlierkamp-Voosen~\nocite{REF:Muhlenbein1993}($1993$). From this
we see that the system actually reaches the optimal state in a
finite time $T$, rather that approaching exponentially.  This is
due to the square-root behavior of the velocity near $\qq=1$ noted
above. $T$ is given by
\begin{equation}
T=(\frac{\pi }{2}+\beta )\frac{{\sqrt{L}}}{G(\phi )}.
\label{EQ:RecomFinite}
\end{equation}
A surprising feature of this result is that $T$ is finite even in the
limit of $\qq(0)\rightarrow 0$, i.e., a random initial population that
has an arbitrarily small chance of having matches in its chromosomes.
Hence, the maximum amount of time $T_{\textit max}$ the population needs to
converge to the optimum is
\begin{equation}
T_{\textit max}={\pi }\frac{{\sqrt{L}}}{G(\phi )}.
\label{EQ:Finite}
\end{equation}

To appreciate this result and the benefit of recombination, it is
helpful to compare it with that of pure enrichment (i.e.,
selection only), given the same initial condition. Starting from
$\qq(0)$, the population distribution initially is a binomial
distribution. Selection keeps chopping off the low tail of this
distribution round by round, and stops when the distribution
contains only the perfect state with $L$ matches. Note that since
here we care about the extreme tail of the distribution,  the
Gaussian approximation is no longer adequate. Based on this
scenario, the evolution time $T$ is determined by the overall
fraction of the population remaining after $T$ rounds, $\phi^T$,
which must be the fraction starting out in the perfect state,
$\qq(0)^L$.  Therefore,
\begin{equation}
{T}= {L} \frac{\ln\,\qq(0)}{\ln\phi},
\label{EQ:SELEX}
\end{equation}
which \textit{diverges} when $\qq(0)\rightarrow 0$. This divergence can
be understood as a result of the initial distribution becoming both
closer to $0$ and narrower in width as $\qq(0)\rightarrow 0$. This
behavior is dramatically different from that for the recombination
case shown in Eq.~(\ref{EQ:Finite}).  Comparing Eq.~(\ref{EQ:SELEX})
with Eq.~(\ref{EQ:RecomFinite}), we see another significant
difference. The evolution time scales differently with $L$
in the two cases; as $\sqrt{L}$ in the case of recombination and as $L$ in
the case of enrichment. This means that the longer the chromosome, the
greater the benefit of recombination.

We have shown that recombination provides a significant
improvement when $\qq(0)$ is very small. In general, it can be
seen by comparing Eq.~(\ref{EQ:SELEX}) with
Eq.~(\ref{EQ:RecomFinite}) that under the conditions when the
evolution takes a longer time [i.e, the worse the initial
condition, the weaker the selection strength (larger $\phi$) or
the longer the chromosome], recombination is highly beneficial.
Under the opposite conditions, recombination is usually still not
worse.  There are cases where recombination does not confer any
benefit, for example if the selection strength is so strong that
the selected population all belongs to the optimal state. However,
when $\qq(0)$ is very close to one, recombination actually takes
longer, since $T$ vanishes linearly in $1-\qq(0)$ without
recombination, and as a square-root, $\sqrt{1-\qq(0)}$ with.  We
shall encounter a similar phenomenon when we discuss the
equilibrium state.  The reason for this is that recombination
repopulates {\em all} the states, whereas pure enrichment does
not.  Since for $p(0)$ very close to 1, the best state is the most
highly populated, recombination can only hurt. Again we must note
in passing that in the case of very strong selection, the results
from our continuum treatment deviate from those of the actual
discrete process.

{\bf Dynamics with homogeneous initial condition:}
So far, we have studied the dynamics of maximal recombination in the
absence of mutation. Now we insert the point mutation process into
the evolutionary dynamics. As a first step, we again assume a
homogeneous initial condition such that each locus has the same
probability of being favorable (i.e., a match) $\qq(0) (\neq 0)$.
Mutation is the final step of the round. Since we only consider
single base mutations, linkage is not introduced in the process. The
mutation process we consider here includes both forward and back
mutations; by itself, it drives the chromosomes toward the maximum
entropy point $L/\ab$ (or $1/\ab$ in terms of $\qq$), which is in
general opposite to the direction of selection. A simple calculation
yields
\begin{equation}
{\qq}_{\textrm{\Pheno}}={\qq}_{\textrm{r}}+
\frac{\MutRate }{\ab -1}(1-\ab {\qq}_{\textrm{r}})
\label{EQ:qmutation}
\end{equation}
Combining this equation with Eq.~(\ref{EQ:qrecomb}), we obtain a
recursion relation for the evolution of $\qq$:
\begin{equation}
{\qq}(t+1)={\qq}(t)+\frac{\MutRate }{\ab -1}(1-\ab {\qq}(t))+
      \frac{G(\phi )}{\sqrt{L}}\sqrt{{\qq}(t)\left(1-{\qq}(t)\right)} ,
\label{EQ:Evol1}
\end{equation}
where we have dropped a correction factor $(1+\MutRate\ab/(\ab-1))$ in the
last term, since $\MutRate \ll 1$.  It is clear that the second term
on the right hand side of the recursion relation is due to mutation,
and the third term to selection and recombination.  When the third
term dominates the second term, the dynamics is essentially the same
as that with no mutation as discussed above.
When $\qq>1/\ab$, the mutational contribution becomes harmful to the
evolution.

The recursion relation Eq.~(\ref{EQ:Evol1}) has an interesting feature.
If we divide Eq.~(\ref{EQ:Evol1}) by $\MutRate$ on both side, we have
\begin{equation}
\frac{{\qq}(t+1)-{\qq}(t)}{\MutRate }=
\frac{1}{\ab -1}(1-\ab {\qq}(t))+
\frac{G(\phi )}{\MutRate \sqrt{L}}\sqrt{{\qq}(t)\left(1-{\qq}(t)\right)}.
\label{EQ:Evol2}
\end{equation}
Eq.~(\ref{EQ:Evol2}) says that the scaled combination ${G(\phi
)}/(\MutRate \sqrt{L})$ is the single control parameter (as opposed
to ${G(\phi)}/{\sqrt{L}}$ in the case of no mutation).  In other
words, if different choices of parameters result in the same
combination ${G(\phi )}/({\MutRate \sqrt{L}})$, the corresponding
dynamics are exactly the same as long as the time scale in each case
is rescaled by its respective mutation rate $\MutRate$. Note that in
Eq.~(\ref{EQ:Evol1}) or Eq.~(\ref{EQ:Evol2}), $\qq(t)$ can go above $1$
when selection is strong; this unrealistic result comes from the
Gaussian approximation to the population distribution that becomes
inaccurate when the population reaches the neighborhood of the
optimal state.  We will further address the error due to the Gaussian
approximation in our discussion of the equilibrium state.

\setlength{\unitlength}{1mm}%
\begin{figure}
\includegraphics*[height=3in]{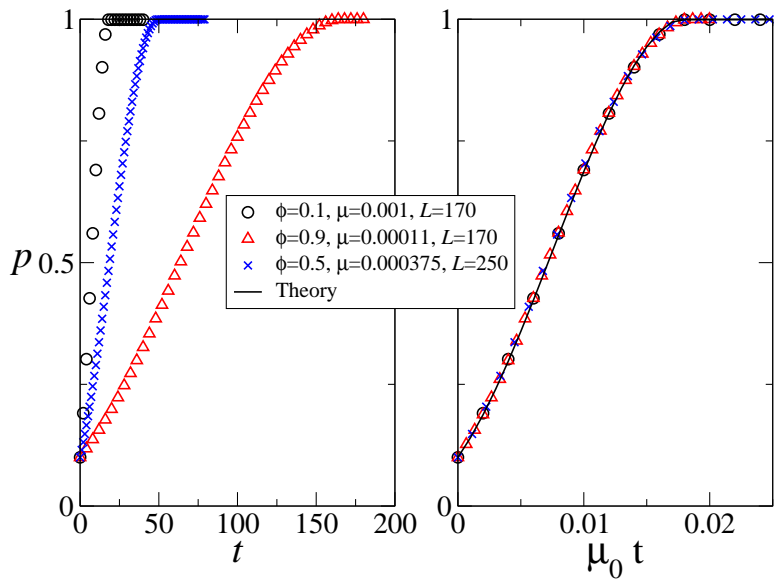}
\caption{\protect \flushing \label{FIG:DataClapsingDynamics}
Comparison of simulation results and theory. The evolutionary
trajectories of three simulations with totally different parameters
[but same scaling variable ${G(\phi )}/({\MutRate \sqrt{L}})$]
collapse into a single curve when time is properly rescaled, and this
curve agrees well with the theoretical result obtained from
Eq.~(\ref{EQ:Evol1}). Here the three simulations all start from a
homogeneous initial condition of $\qq(0)=0.1$ and population size
$N=10^4$.}
\end{figure}

As a test of our theory, Fig.~(\ref{FIG:DataClapsingDynamics}) shows
that the theoretical predictions, including the scaling with $G(\phi
)/(\MutRate \sqrt{L})$, agree extremely well with simulation data.
This indicates that the finite-population effect in this landscape is
insignificant, and the Gaussian approximation to the binomial
distribution is appropriate for sequences of long length $L$.

When selection is weak so that the change in $\qq$ in each round is
small, the evolution equation~(\ref{EQ:Evol1}) can again be
accurately approximated by its continuous time version,
\begin{equation}
\frac{d\qq}{dt^{\prime }}=(1-\ab \qq) +C\sqrt{\qq(1-\qq)},
\label{EQ:ContinuousVer0}
\end{equation}
where $t^{\prime }\equiv (\MutRate/(\ab-1))\, t$ and
$C\equiv (\ab-1)G(\phi)/(\MutRate \sqrt{L})$.  The explicit solution of
this equation can be found in Appendix~\ref{APP:ExplicitSolution}.

An explicit analytical comparison of the evolutionary performance
between the evolutionary protocol with and without recombination is
not available. Both recombination and mutation can serve to generate
diversity, but the greater benefit of recombination is due to that
facts that (a)~recombination breaks up the linkage introduced by
selection much more effectively than mutation does; breaking of
linkage helps broaden the distribution (as selection narrows the
distribution) and in turn facilitates more efficient future
selection, hence speeding up the evolution, and (b)~recombination
keeps the mean of the population unchanged, whereas mutation goes
against selection (once the population goes beyond the maximum
entropy point). Therefore, as has been recognized for a long time,
recombination is able to generate variety without the excessive
baggage of deleterious mutations.

{\bf Dynamics with inhomogeneous initial condition:}
In the previous discussion, we assumed a simple homogeneous initial
condition. An immediate question concerns what happens with a
inhomogeneous initial condition, i.e. would different loci
synchronize with each other after a short transient period or would
they go their separate ways and only meet at the end of the process?
To answer this question, we choose a linkage-free initial condition
where half of the loci have probability $\qq_{1}(0)$ of being a match,
and the other half have probability $\qq_{2}(0)$ of being a match.
Assuming again that $L\gg 1$, a similar derivation to the one
presented above produces
\begin{eqnarray}
\qq_{1}(t+1) & = & \qq_{1}(t)+\frac{\MutRate (1-\ab \qq_{1}(t))}{\ab -1}+\frac{G(\phi )}{\sqrt{L/2}}\frac{{\qq}_{1}(t)(1-{\qq}_{1}(t))}{\sqrt{{\qq}_{1}(t)(1-{\qq}_{1}(t))+{\qq}_{2}(t)(1-{\qq}_{2}(t))}}
\nonumber\\
\qq_{2}(t+1) & = & \qq_{2}(t)+\frac{\MutRate (1-\ab \qq_{2}(t))}{\ab -1}+\frac{G(\phi )}{\sqrt{L/2}}\frac{{\qq}_{2}(t)(1-{\qq}_{2}(t))}{\sqrt{{\qq}_{1}(t)(1-{\qq}_{1}(t))+{\qq}_{2}(t)(1-{\qq}_{2}(t))}}
\label{EQ:MixedDynamics}
\end{eqnarray}

\begin{figure}
\includegraphics*[  height=2.5in]{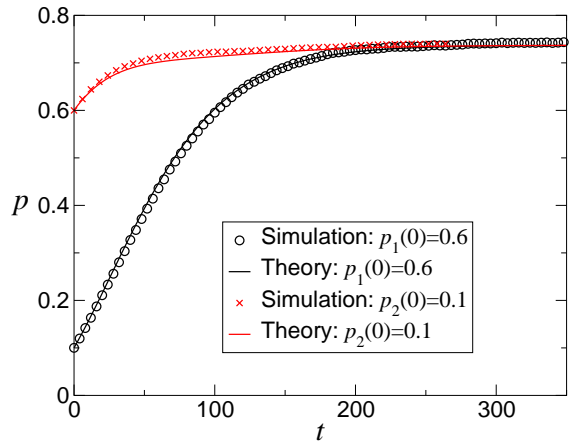}
\caption{\protect \flushing Evolution of individual match
probability $\qq$ with inhomogeneous initial condition. The initial
condition is chosen so that half of the loci have initial match
probability of $\qq_{1}(0)=.1$, while the remaining half has
$\qq_{2}(0)=.6$. The two simulation curves track the two different
groups of loci. The solid curves are theoretical results from
Eq.~(\ref{EQ:MixedDynamics}).
Simulations use
the following parameters: $L=170$, $\phi=.9$, $\MutRate=.01$ and $N=10^4$.
\label{FIG:MeanEvo_MixedInit_MaxRecom_Mutation} }
\end{figure}

\begin{figure}
\includegraphics*[ height=2.5in]{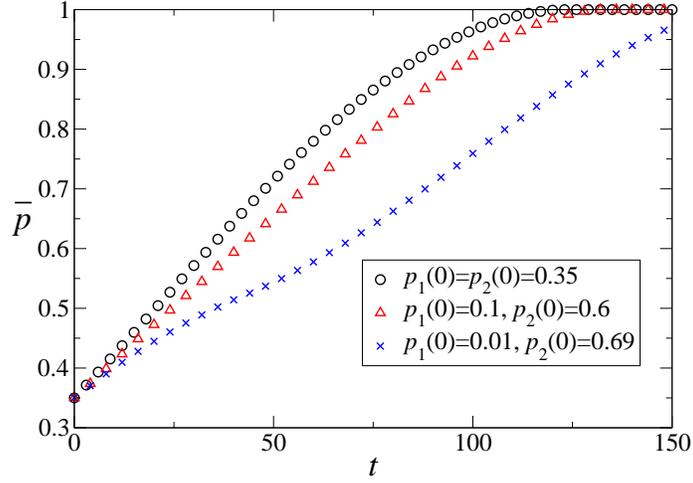}
\caption{\protect \flushing Comparison of simulation results for the
average individual match probability $\overline \qq$ for different
initial conditions, in the absence of mutation. $\overline \qq$ is the
individual match probability $\qq$ averaged over loci. Here $L=170$,
$\phi=.9$ and $N=10^4$.
\label{FIG:MeanEvo_MixedInit_MaxRecom_NOMutation} }
\end{figure}

Fig.~(\ref{FIG:MeanEvo_MixedInit_MaxRecom_Mutation}) compares these
results with a evolutionary trajectory with inhomogeneous initial
conditions. It is clear that
different locus go their own ways. %
If mutation is ignored, one find that the relative changes in the
individual match probabilities are
\begin{equation}
\frac{\qq_{1}(t+1)-\qq_{1}(t)}{\qq_{2}(t+1)-\qq_{2}(t)}=\frac{\qq_{1}(t)(1-\qq_{1}(t))}{\qq_{2}(t)(1-\qq_{2}(t))},
\end{equation}
 i.e., proportional to the ratio of the variance on each locus. In other words, the
selection works on variance; different locus with different $\qq$
experience different selection pressures depending on their
contribution to variance. This means that in the case of
inhomogeneous initial conditions, the evolution process will be
dominated by whichever loci have very small initial $\qq(0)$; see
Fig.~(\ref{FIG:MeanEvo_MixedInit_MaxRecom_NOMutation}) for an
explicit demonstration of this conclusion.

{\bf Dynamics with clonal initial condition:}
In the above theoretical analysis we have
assumed an initial condition which already has all the favorable
alleles available in the population, and mutation was relatively
unimportant.  In many situations, mutations are absolutely
necessary for the system to find the optimal state.  This scenario
could happen for a system prepared with a clone-like initial
condition with some loci lacking the beneficial alleles.  In this
section we focus on this case and study the effect of different
mutation rates on the evolutionary dynamics. We start from the
simplest case which is a clonal initial condition with no beneficial
allele at any locus (i.e., $\qq(0)=0$).  For this simple case, the
obviously best strategy is to subject the system to maximal mutation
rate $\MutRate = (\ab-1)/\ab$ until the system reaches $\qq = 1/\ab$,
and then to stop mutation altogether. However, this naive strategy
does not apply to other clonal initial conditions which are
inhomogeneous; here it is not a priori clear when one should turn off
mutation, since we have access only to measurements of the full
phenotype.

\begin{figure}
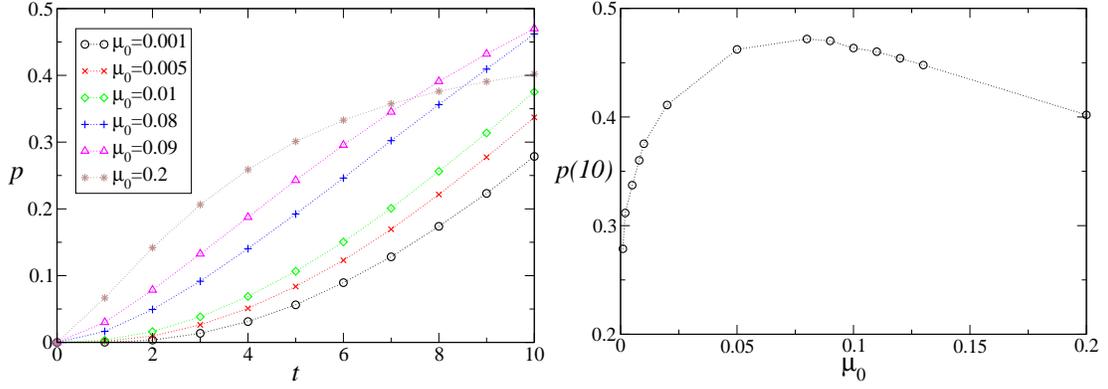

\includegraphics*[ height=2. in]{DiffEvoTraj.eps}
\includegraphics*[ height=2. in]{MutationPerformance.eps}
\caption{\protect \flushing Simulation results
 of different mutation rates given
a clone  initial condition of $\qq(0)=0$. (a)~Evolutionary trajectories for different choices of mutation rate $\MutRate$. (b)~Comparison of individual match
probability at the tenth generation $\qq(10)$ for different mutation
rates. The "optimal" mutation rate is around $\MutRate =.08$
under this condition. Here $L=170$, $\phi=.1$ and $N=10^4$.
\label{FIG:MutationPerformance} }
\end{figure}

\begin{figure}
\includegraphics*[ height=2.5 in]{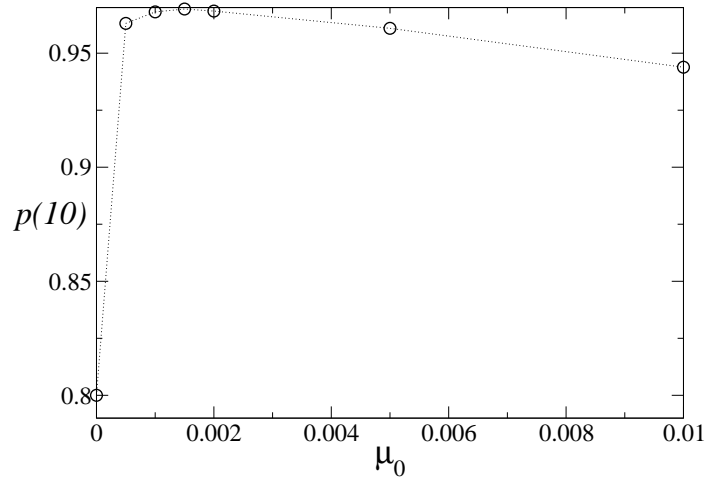}
\caption{\protect \flushing Simulation results for different mutation
rates given an inhomogeneous clonal initial condition. Initially, all
the member of the population have the same haplotype; $1/5$ of the
loci are occupied by mismatches and the rest by matches. Here
$L=170$, $\phi=.1$ and $N=5\times 10^5$.
\label{FIG:MutationPerformanceClone2} }
\end{figure}

Assuming that mutation rate is fixed, we can gauge the performance of
different choices of mutation rate by measuring $\qq$ after a given
number of generations. Fig.~(\ref{FIG:MutationPerformance}) shows a
comparison of simulations starting at $\qq(0)=0$ with different
mutation rate using this criterion. It is clear that there is a
``optimal'' mutation rate. Of course, the "optimal" mutation rate
depends when evolution is terminated.
In the more common situation, a majority of the loci are already
occupied by beneficial alleles, and for the rest, mutation is
required to create the beneficial allele. In this case as well, we
can see the existence of an ``optimal'' mutation rate, as in
Fig.~(\ref{FIG:MutationPerformanceClone2}).
\setlength{\unitlength}{1mm}%

In conclusion, mutation is necessary when there is limited diversity
in the initial library. Even in this case, mutation is only helpful
in the short term dynamics, but harmful to the equilibrium state. In
a realistic experiment, since breeding usually proceeds for a only
few generations there would exist an ``optimal'' mutation rate.

{\bf Equilibrium state of maximal recombination:}
Having studied the dynamical behavior of the model in the previous
sections, we now turn to the study of its equilibrium properties,
focusing on the mean number of matches in equilibrium. This
quantity plays the role here that genetic load does in normal
evolution problems. Strictly speaking, genetic load is the
difference in fitness between the equilibrium population and the
optimal fitness state. In our case, the optimal state has fitness
value $1$, and the mean fitness of the population is simply
$\phi$, so the difference is trivially $1-\phi$. However, since
the purpose of the experiments is to maximize the phenotypic
value, i.e., the number of matches, the mean number of matches is
a reasonable way to characterize the equilibrium state, and the
``cost'' of mutations. Going back to the language of DNA protein
binding, what we are using to characterize the equilibrium is the
mean binding affinity of the DNA sequence.

Under the Gaussian approximation,
the equilibrium value $\qE$ can be found by setting $\qq(t+1)=\qq(t)$ in
Eq.(~\ref{EQ:Evol1}), yielding
\begin{equation}
\frac{\qE(1-\qE)}{(\ab \qE-1)^{2}}=\frac{1}{(\ab
  -1)^{2}}\left(\frac{\MutRate \sqrt{L}}{G(\phi
  )}\right)^{2}
\label{EQ:EquilibriumP}
\end{equation}
\setlength{\unitlength}{1mm}%
\begin{figure}
\includegraphics*[  height=2.5in]{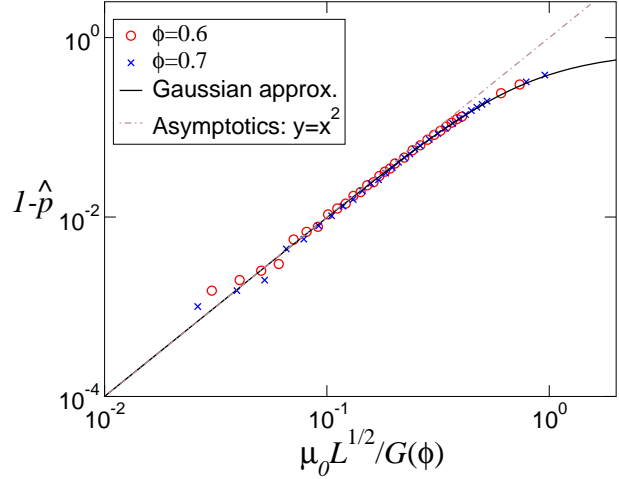}
\caption{\protect \flushing Equilibrium value of individual match
probability $\qq$ as function of ${\MutRate \sqrt{L}}/{G(\phi )}$.  The
$y=x^2$ curve denotes the asymptotic behavior of $\qq$ in strong
selection (low mutation) resulting from the Gaussian approximation,
Eq.~(\ref{EQ:EquilibriumPStrong}). For simulation results: $L=170$,
$\MutRate=.01$, $N=10^4$.
\label{FIG:EquiPositionScaling}}
\end{figure}
As shown in Fig.~(\ref{FIG:EquiPositionScaling}), for weak and
intermediate selection strength this result agrees with the
simulation data. For the parameter regime of strong selection (or
very small mutation rate), ${\MutRate \sqrt{L} }/{G(\phi )}$ is no
longer a good scaling variable, and the Gaussian approximation
begins to deviate from the exact result.

For weak selection (i.e., $({\MutRate \sqrt{L}}/{G(\phi )})^{2}\gg
1$) the equilibrium probability is near $1/\ab$. We have
\begin{equation}
\qE-\frac{1}{\ab}=\frac{(\ab -1)^{\frac{3}{2}}} {\ab ^2} \frac{G(\phi
)}{\MutRate \sqrt{L}}
\label{EQ:EquilWeak}
\end{equation}
When selection is relatively strong, i.e., when $({\MutRate\sqrt{L} }/{G(\phi )})^{2}\ll 1$,
the equilibrium position is
\begin{equation}
1-\qE=\left(\frac{\MutRate\sqrt{L} }{G(\phi
  )}\right)^{2}
\label{EQ:EquilibriumPStrong}
\end{equation}
As shown in Fig.~(\ref{FIG:EquiPositionScaling}), for the
parameters employed there, this power law is accurate in the
region of $0.1\le ({\MutRate\sqrt{L} }/{G(\phi )})^{2} \le 1$. ( A
more precise statement about the lower cutoff for the validity of
Eq.~(\ref{EQ:EquilibriumPStrong}) is $ \MutRate L > \ln
\,\phi^{-1} $; see discussion below). This $\MutRate^2$ dependence
has also been derived by
Kondrashov~\nocite{REF:Kondrashov1995}($1995$)  for a different
type of dynamical selection scheme.

It is well know that in a fixed smooth landscape, the mutational
load is $\MutRate L $ in the strong selection limit. We see from
the above that for our breeding problem, strong selection, at
least in the Gaussian approximation, produces a load that scales
with $\MutRate ^{2}$ rather than the usual $\MutRate$. The reason
for this scaling is easy to understand: With recombination, the
change due to selection of $\qq L$, the number of matches, is
proportional to the width of the distribution
$\sqrt{L\qq(1-\qq)}\approx \sqrt{L(1-\qq)}$ (which is independent
of the Gaussian approximation), and mutation reduces $\qq L$ by
$-\MutRate L$. Balancing the two effects we arrive at the
$\MutRate ^{2}$ scaling. This simple argument shows that the
$\MutRate ^{2}$ law is a generic result for the genetic load when
recombination is at work (which happens whenever the selected
population still occupies a number of different states). When
selection strength is sufficiently strong that the selected
population lies almost entirely in the optimal state,
recombination becomes ineffective as there is no diversity within
the selected population. As a result [as shown in
Fig.~(\ref{FIG:EquiPositionScaling})], the Gaussian result is no
longer valid. In fact, as we shall show below, in this case the
usual mutational load of O($\MutRate$) takes over.

For small mutation rates, as we shall see, the mean mismatch number is small
at equilibrium, except for very weak selection.  This is in contradistinction
to the dynamic case, where we focused on the case where there were a large
number (order $L$) of initial mismatches.  There, for most of the time,
the Gaussian approximation is quite adequate.  In equilibrium, however,
this is generically not the case, and a more careful treatment is warranted.

To study this issue in more detail, we can make use of a
more powerful approximation scheme that accurately covers both
the strong selection and extremely strong selection cases so as to
find the equilibrium position. Since $L\qq\sim L$, we shift our focus
to the mismatch probability $\pp\equiv1-\qq\sim 0$, and, before
selection, the population distribution in terms of mismatch should
follow a Poisson distribution
$P_{k}=\PP^{k}e^{-\PP}/{k!}$, where
$\PP\equiv L\pp$.  Call $C_{k_{0}}$ the cumulative probability
of being in a state with $k\leq {k_{0}}$. Selection results in
\begin{equation}
\phi =C_{k_{0}-1}+\alpha P_{k_{0}},
\label{EQ:EquilibriumPoisson}
\end{equation}
where $k_{0}$ is the threshold, which is determined by the condition
that $C_{k_{0}-1} < \phi \leq C_{k_{0}}$. $\alpha \leq 1$ counts the
partial selection on the members of population with $k_{0}$
mismatches.

\begin{figure}
\includegraphics*[  height=2.5in]{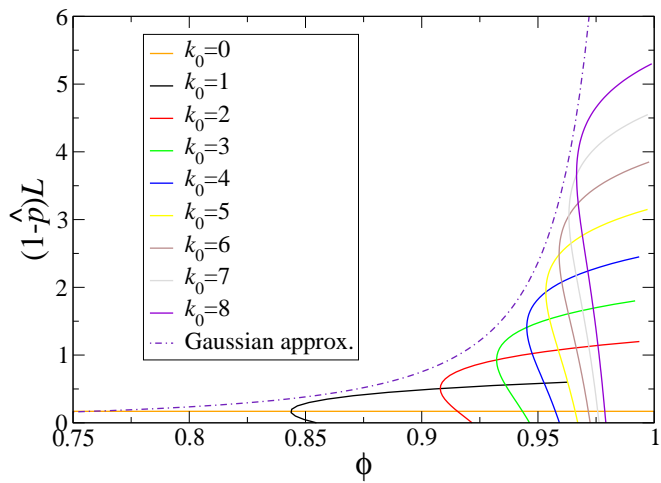}
\caption{\protect\flushing Theoretical evaluation of equilibrium
value of mean mismatches: comparison of Gaussian approximation and
Poisson approximation.
Curves corresponding to different $k_0$ values are trial solutions of
Eq.~(\ref{EQ:EquilibriumPoissonResult}) with different choices of
threshold $k_0$. The envelop of the trial solutions is the physical
solution. For comparison, the solution given by Gaussian
approximation is also plotted. Here $L=170$, $\MutRate=.001$.
\label{FIG:Poisson}}
\end{figure}

The result of selection, recombination and mutation are
\begin{eqnarray}
\pp_{\textrm{r}}&=&\frac{1}{L\phi }\left(\sum
_{k=0}^{k_{0}-1}kP_{k}+\alpha
k_{0}P_{k_{0}}\right),
\label{EQ:EquilibriumPoisson15}
\\
\pp_{\textrm{\Pheno}}&=& \pp_{\textrm{r}} + \MutRate - \frac{\ab}{\ab-1}\MutRate \pp_{\textrm{r}},
\label{EQ:EquilibriumPoisson2}
\end{eqnarray}
where $\pp_{\textrm{r}}$ ($\pp_{\textrm{\Pheno}}$) is the individual mismatch
probability right after recombination (mutation).  The last term is
negligible as it arises from extremely rare compensatory beneficial
mutations. Combining Eqs.~(\ref{EQ:EquilibriumPoisson}, \ref{EQ:EquilibriumPoisson15},
\ref{EQ:EquilibriumPoisson2}), we arrive at the following equation
that determines the equilibrium average mismatch number $\PE$:
\begin{equation}
\phi =\frac{e^{-\PE}}{k_{0}+\Mut-\PE}\sum
_{k=0}^{k_{0}-1}\frac{k_{0}-k}{k!}{\PE}^{k},
\label{EQ:EquilibriumPoissonResult}
\end{equation}
where $\Mut \equiv \MutRate L$ is the genomic mutation rate. The
physical $\PE(\phi )$ curve is given by the envelope
of the various $\pE(\phi ,k_{0})$ for different
$k_{0}$'s, and is piecewise analytic. Fig.~(\ref{FIG:Poisson}) shows
the difference between this solution and the solution given by the
Gaussian approximation.  Even though the Gaussian approximation
has generally the right trend, for this small mutation rate the
Gaussian approximation is numerically off by a large percentage, except
for extremely weak selection. Also, for $\phi <e^{-\Mut }$,  selection only
preserves the optimal state at equilibrium, recombination stops
working, and Eq.~(\ref{EQ:EquilibriumPoissonResult}) produces
$\PE=L-L\qE=\Mut$, the conventional mutational load,
which is the result that is unobtainable from the Gaussian
approximation.
\setlength{\unitlength}{1mm}%

\begin{figure}
\includegraphics*[  height=2.5in]{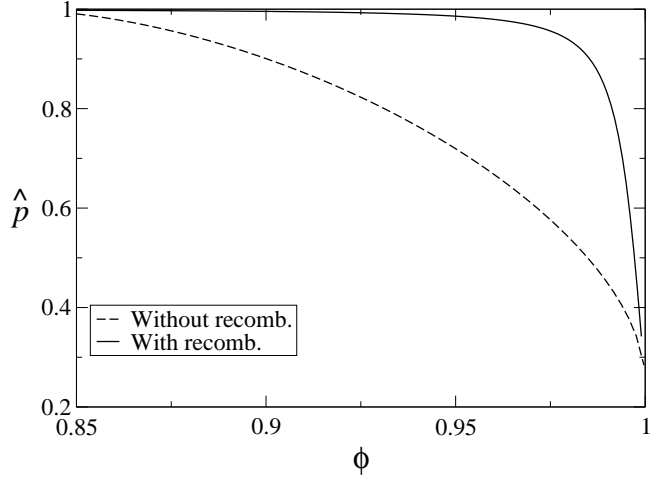}
\caption{\protect \flushing Comparison of equilibrium positions for
two evolutionary protocols: with recombination
[From Eq.~(\ref{EQ:EquilibriumP})] and without
recombination [From Eq.~(\ref{EQ:nosex})].
It is clear that the benefit of recombination is most
significant for intermediate selection strength.  Here $L=170$ and $\MutRate
=.001$. \label{FIG:EquilibriumPostionComparison}}
\end{figure}

Now we come back to the question of how much of an improvement is
conferred by recombination.
The easiest way to approach the question is through a graph, Fig.
(\ref{FIG:EquilibriumPostionComparison}), directly comparing the
equilibrium state with and without recombination, for the same
value of $\MutRate$. To do this, we use the analysis for the no
recombination case from Cohen and
Kessler~\nocite{REF:Kessler2002}($2003$):
\begin{equation}
-\frac{\ln \phi}{\Mut}=
\frac{1}{\ab-1}+\frac{\ab-2}{\ab-1}\qE_{\rm A}
-\sqrt{\frac{4}{\ab-1}\qE_{\rm A}(1-\qE_{\rm A})}.
\label{EQ:nosex}
\end{equation}
Here $\qE_{\rm A}$ denotes the equilibrium value of total number of
matches divided by chromosome length $L$ for the no-recombination
case.  The first thing to notice is that at the two extreme limits of
maximal selection and no selection, the two approaches yield the same
result. The curves are nevertheless very different.  In order to
understand this difference better, we first focus on the case of
intermediate selection.  We re-express Eq.
(\ref{EQ:EquilibriumPStrong}) to simplify the comparison with Eq.
(\ref{EQ:nosex}).  Using the small $\phi$ limit of $G$, we obtain
\begin{equation}
-\frac{\ln \phi}{\Mut} = \frac{\MutRate }{2(1-\qE)}
\end{equation}
Without recombination, $\qE$ is an order $1$ function of
$\chi\equiv -\ln \phi /\Mut$, whereas with recombination, the
relation involves $\MutRate$ as well as $\chi$.  Thus, with
recombination, the only way $\chi$ can be of order 1 is if $1-\qE$
is of order $\MutRate$, i.e. small.  Thus, the recombination curve
remains near $\qE=1$ until $\phi$ approaches 1, at which point it
takes a sharp dive.  Thus, recombination leaves the equilibrium
state much less sensitive to selection, and very close to the
optimal state, unless the selection is very weak.

It is also interesting to compare the two problems in the weak selection
limit. The no-recombination result Eq. (\ref{EQ:nosex}) implies that
\begin{equation}
\qE_{\rm A}-\frac{1}{\ab} =\frac{2(\ab-1)}{\ab^{3/2}}\sqrt{\frac{(1-\phi)}{\Mut}}
\end{equation}
so that $\qE_{\rm A}$ has a square root singularity at $\phi=1$. On the other
hand, the recombination result Eq. (\ref{EQ:EquilWeak}) for $\phi \approx 1$
reads
\begin{equation}
\qE-\frac{1}{\ab} = \frac{(\ab-1)^{3/2}
}{\ab^2}\frac{(1-\phi)\sqrt{-2\ln(1-\phi)} }{\MutRate \sqrt{L}}
\end{equation}
so that, modulo the weak logarithmic singularity, $\qE$ is
essentially linear at $\phi=1$.  However, the slope is very large, of
order $1/\sqrt{\MutRate \Mut}$, a factor of $1/\sqrt\MutRate$ larger
than the $1/\sqrt{\Mut}$ coefficient of the square-root singularity of
the no-recombination result. Thus, we have the anomalous result that
for sufficiently weak selection, recombination actually makes the
equilibrium state worse.  We can calculate the cross-over point below
which recombination loses its superiority; it is given by $1-\phi
\propto (\MutRate/\ln \MutRate)^{1/2}$, which is small for small
$\MutRate$.

\setlength{\unitlength}{1mm}%

{\bf Finite population effect:}
\setlength{\unitlength}{1mm}%
\begin{figure}
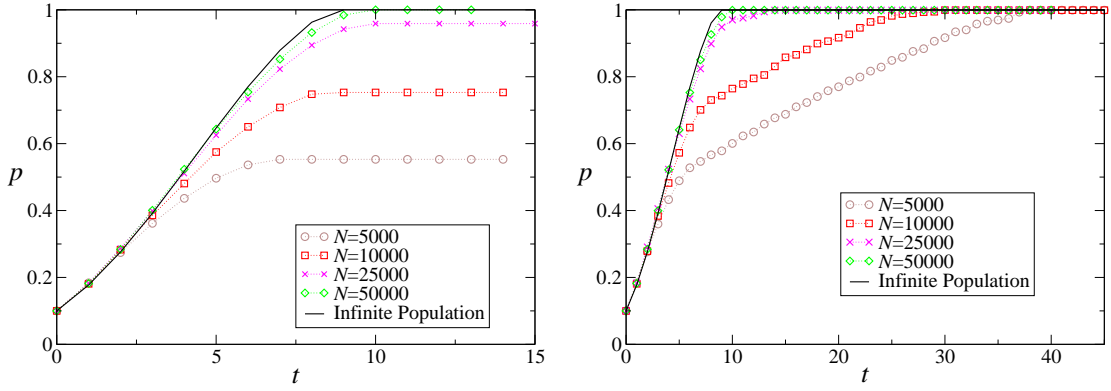

\includegraphics*[  height=2in]{Mu.00.eps}\hspace{.3cm}
\includegraphics*[  height=2in]{Mu.001.eps}
\caption{\protect \flushing  Finite Population effect on the dynamics
and equilibrium of maximal recombination: Simulation results for the
case of (a)No mutation ($\MutRate=0$) and (b) Weak mutation
($\MutRate=0.001$). With mutation, two regimes exist: a)~at the
beginning recombination dominates evolution, as in the mutation-free
case and (b)~at later stages mutations are essential, as they are
needed to find lost favorable alleles. Here evolution is constrained
by the rate for mutation to find beneficial alleles, thus slower.
Here $L=170$, $\phi=.1$. Infinite population trajectories are
theoretical results taken from Eq.~(\ref{EQ:Evol1}).
\label{FIG:Mu00}}
\end{figure}
In the previous sections, we have assumed an infinite population, and
comparison with simulations have shown that this approximation is
appropriate most of the time. However, under the experimentally
relevant situations, where very strong selection and very low
mutation rate are employed, finite population effects can show up and
be potentially important. As an extreme example, if we choose the
best member from the population, i.e, $N\phi=1$, then recombination
has nothing to work with and offers no benefit. In this section we
discuss the population size above which the evolutionary dynamics and
equilibrium can essentially be deemed the same as those of infinite
populations. To rid ourselves of the impact of initial conditions, we
assume an initially diversified population. Two population sizes are
relevant for this problem, the total population size $N$ at the
beginning of each generation, and the selected population size
$N\phi$. We will focus on the case most often encountered in
\textit{in vitro} evolution experiments, namely $N$ very large, but
$N\phi$ is rather small. In the following analysis, we set $\phi$
constant, and compare the evolutionary trajectory for different
population sizes $N$. We start again from the simplest case of no
point mutation, shown in Fig.~(\ref{FIG:Mu00}a). It can be seen that
in general the smaller the population size, the worse the
evolutionary efficiency and the equilibrium state. Under these
conditions, the finite population effect comes from the genetic
association between loci that results in the loss of favorable
alleles from the entire population when the population is subject to
selection.

We can estimate the population size $N_{0}$ above which the
population dynamic approaches infinite-population results by
estimating the average number of lost favorable alleles at all loci
in the selected population.  As the individual match probability $\qq$
is lowest at the beginning of evolution, if there is any loss of
favorable allele, it is most likely to happen at the first round of
selection, hence we focus on the first round of selection and
recombination. Right after the first round of selection, the
probability for a specific locus to lose the favorable allele is
$(1-\qq(1))^{N\phi}$. Therefore the average number of loci losing the
favorable allele is $L(1-\qq)^{N\phi }$. The opposite way of saying
this is that for the population to retain favorable alleles at all
loci after the first round of selection, we require $L(1-\qq)^{N_0\phi
}<1$. This gives the following estimate of $N_{0}$:
\begin{equation}
N_{0}  =  \frac{1}{\phi }\frac{-\ln L}{\ln (1-\qq(1))},
\label{EQ:FiniteSize}
\end{equation}
where  $\qq(1) = \qq(0)+G(\phi )\sqrt{\qq(0)(1-\qq(0))/L}$. If in the first
round the above criterion is met, then it is more unlikely for the
population to lose any favorable alleles any at successive rounds
when subject to selection, as $\qq(t)$ increases in later rounds and
$L(1-\qq)^{N\phi }$ becomes increasingly smaller than $1$. Comparison
with simulation results shows that Eq.~(\ref{EQ:FiniteSize}) gives a
reasonable estimate (at least correct in order of magnitude).


When a (weak) point mutation is involved, the dynamics can be
separated into two regimes: (a)~recombination dominated evolution,
where for loci that have a favorable allele in them, recombination
helps that favorable loci to spread to the whole population, as in
the mutation-free case,  and (b)~mutation/recombination evolution,
where loci that have lost their favorable alleles due to selection
need mutations to find them again. The dynamics in this regime is
significantly slower, as seen in  Fig.~(\ref{FIG:Mu00}b).  In this
case, Eq.~(\ref{EQ:FiniteSize}) still identifies the smallest
population size $N_0$ that behaves more or less like an infinite
population. As a side note, the flip side of the finite population
issue is that there exists an optimal selection pressure for a
particular population size; an estimate of this optimal $\phi $ can
be found from the condition $L(1-\qq)^{N\phi }<1$.


The estimate of $N_{0}$ given $\phi $ (or equivalently the optimal
$\phi_0 $ given $N$) has practical implications. It gives a
population size that is just enough for the evolution to proceed in
its fastest rate (given a selection strength). Since in real-life
experiments selection routinely involves screening the population,
which is costly and time-consuming, a smaller while still equally
effective population size would be helpful. Of course, the above
arguments assumed a random initial library. If one starts from a
population of a single clone, then as already explained above, even
at the beginning mutation is absolutely necessary to generate
diversity. For this case it is more difficult to directly estimate
the finite population effects.\begin{center}\end{center}

\begin{center}
\section*{DISCUSSION}
\end{center}
In this work we proposed and studied a simplified model of DNA
shuffling, a very important evolutionary protocol for directed
evolution. We investigated this model from a population genetics'
point of view, as a multi-locus evolutionary model incorporating both
recombination and point mutation and subject to dynamical truncation
selection. Our specific recombination scheme, as an extreme limit of
multi-parent multi-crossover genetic mixing employed in DNA
shuffling, enables us to pursue analytical results for the dynamical
and equilibrium features. To summarize, we derived the recursion
relations that completely characterize the evolutionary process,
which shows explicitly how recombination helps to speed up the
evolution. Assuming a large number of loci, we found that selection
and mutation affect the evolutionary process only through a
combination of their respective parameters. When the evolution is
relatively slow so that the discrete recursion relation can be
approximated by a continuous differential equation, we could solve
for the evolutionary trajectory, and prove in special cases that it
is indeed faster than that for the case of no-recombination. We also
investigated the equilibrium properties of the system, and found that
the genetic load when recombination is in effect has a different
scaling form than usual, as long as the selection is not super
strong. In addition, we discussed the finite-size effect under
conditions relevant to the shuffling experiments, and estimated the
minimal population size above which the population behaves as if an
infinite population.

We have been focusing on a model which is directly relevant only
under very specific conditions, but our results are actually
applicable to broader situations. Let us first discuss
selection:(a)~For simplicity, we employed a dynamical truncation
selection. In the case of DNA sequence evolution via binding to
protein, selection is ``smoother'' [roughly speaking, the step
corners are rounded~\nocite{VonhippelB86,
GerlandH02}(V{\scriptsize{ON}} H{\scriptsize{IPPEL}} and
B{\scriptsize{ERG}} $1986$; G{\scriptsize{ERLAND}} and
H{\scriptsize{WA}} $2002$)]. This simplification does not cause
any substantial difference in our analytical results as long as
the binding affinity is large (i.e, the step is steep enough).
(b)~Our approach and results are also applicable to the type of
dynamical ({\it soft}) selection studied by
Kondrashov~\nocite{REF:Kondrashov1995}($1995$). There, the
selection is implemented as a function $W(X)$, where
$X=(\Pheno-\overline{\Pheno})/\sigma$, i.e., the phenotypic
difference between the phenotype and the mean $\overline \Pheno$,
divided by the standard deviation $\sigma$ of the phenotypic
distribution. The effect of this type of selection on a population
with a Gaussian distribution is
\begin{equation}
\Pheno_{\rm s}-\Pheno = \sigma \delta, \label{EQ:Kon}
\end{equation}
where $\delta$ is a quantity solely characterized by the selection
function $W(X)$ and does not depend on the population phenotypic
distribution [See Eqs.($7$) and ($14$) of
Kondrashov~\nocite{REF:Kondrashov1995}($1995$)]. It is evident
that Eq.~(\ref{EQ:Kon}) is analogous to Eq.~(\ref{EQ:Selection})
except that one needs to replace $G(\phi)$ with $\delta$.
Therefore, when one exchange $G(\phi)$ with $\delta$, all our
results under the Gaussian approximation remain valid, including
various scaling arguments and solutions to the continuous
evolutionary equations.

Now we move to recombination. One immediate question is what would
change when the recombination scheme is more realistic. A
straightforward generalization of the maximal recombination toward
the realistic situation is that we can cut each DNA sequence into
segments of equal length, with the maximum recombination as the
extreme case of length equal to $1$. Computer simulation shows
that in general the longer the segment length, the worse the
performance. For a general segment length, one can derive a
hierarchy of recursion relations that relate the cumulants of the
population distribution at the current generation to those of
previous generation, an approach that has been applied to a
multi-locus system on a fixed landscapes~\nocite{REF:Barton1991,
REF:Buger2000}(B{\scriptsize{ARTON}} and T{\scriptsize{URELLI}}
 $1991$; B{\scriptsize{\"{U}GER}} $2000$). How to close the chain
of recursion relations remain to be studied. The difficulty is
associated with the remaining linkage within each segment. In any
event,  the maximal recombination model can serve as a theoretical
upper limit which is qualitatively correct and even reasonably
accurate quantitatively.

Our study has been guided by the {\it in vitro} evolution process,
where the intensity of recombination and mutation are both
prescribed by the experimenter. This might be not be the case in a
more natural setting, where the evolution of recombination
(mutation) itself can be an important aspect of the
problem~\nocite{FeldmanOC1996}({\it e.g.}, F{\scriptsize{ELDMAN}}
{\it et.al.} $1996$). Therefore our model does not incorporate a
modifier gene that explicitly controls recombination rate. In such
a modifier approach, the recombination rate can itself evolve
because the modifier gene is under indirect selection due to its
association with other genes on the same chromosome under direct
selection~\nocite{FeldmanOC1996}({\it e.g.},
F{\scriptsize{ELDMAN}} {\it et.al.} $1996$).

Finally, we believe that our model may be relevant to some
examples of natural evolution: (a) dynamical truncation selection
can also exist in nature, for example the mutual selection in a
host-parasite system and (b) recombination of multi-parents,
multi-crossover type does exist in nature: certain RNA viruses
have multiple segments in their genome, and when multiple viruses
infect the same host and new virus particles are made,
recombination of the type we study can
happen~\nocite{Chao1992}(C{\scriptsize{HAO}} $1988$). With the
ease of analytical treatment of this model, we hope it can serve
as a theoretical testing ground where general statements about the
interplay among various aspects such as recombination, mutation
and selection, can be tried out.

\begin{center}
{The acknowledgements} \end{center}
 We would like to give special
thanks to L. Chao for stimulating discussions. We also acknowledge
helpful interactions with S. P. Otto, A. Poon, W. P. C. Stemmer
and J. Widom. This work has been partially funded by the NSF
sponsored Center for Theoretical Biological Physics (grants \#
PHY-0216576 and 0225630).

\bibliographystyle{authordate1}
\bibliography{reference}

\appendix
\begin{center}\end{center}
\section{Explicit solution of the evolution equation}
\label{APP:ExplicitSolution}
 Assuming that at $t=0$
the individual match probability is $\qq(0)$, the explicit
solution for the continuous version of Eq.~(\ref{EQ:Evol2})(which
is a good approximation when change of $\qq$ in between
consecutive rounds is small, i.e., when selection is not very
strong) is
\begin{equation}
\frac{d\qq}{dt^{\prime }}=(1-\ab
\qq)+C\sqrt{\qq(1-\qq)},
\label{EQ:ContinuousVer}
\end{equation}
where $t^{\prime }\equiv (\MutRate
/(\ab -1))t$ and $C\equiv (\ab -1)G(\phi )/(\MutRate \sqrt{L})$.  For
most parameter ranges the righthand side is positive, therefore $\qq$
keeps increasing.

The general solution of Eq.~(\ref{EQ:ContinuousVer}) is:
 \begin{eqnarray}
& & \frac{\ab ^{2}+C^{2}}{2}t^\prime=-\frac{\ab }{2}\ln
 \frac{\left\vert1-\ab \qq+C\sqrt{\qq(1-\qq)}\right\vert}{\left\vert 1-\ab
 \qq(0)+C\sqrt{\qq(0)(1-\qq(0))}\right\vert}+
 \frac{C(1-\ab/2)}{\sqrt{\Delta }}
\nonumber \\
& & \qquad\qquad\times\left[\ln
 \frac{\left\vert2\sqrt{(1-\qq)/\qq}+{C}-\sqrt{\Delta
 }\right\vert}{2\sqrt{(1-\qq)/\qq}+{C}+\sqrt{\Delta }}-\ln
 \frac{\left\vert2\sqrt{(1-\qq(0))/\qq(0)}+{C}-\sqrt{\Delta
 }\right\vert}{2\sqrt{(1-\qq(0))/\qq(0)}+{C}+\sqrt{\Delta
 }}\right]\nonumber \\
& & \qquad\qquad-C\left(\tan^{-1} \sqrt{\frac{1-\qq}{\qq}}-\tan^{-1} \sqrt{\frac{1-\qq(0)}{\qq(0)}}\right),
\end{eqnarray}
 where $\Delta \equiv C^{2}-4(1-\ab )$. In the absence of mutation the above equation has been solved, the solution a sinusoidal function (see Eq.~(\ref{EQ:SimShufflingSolution}). The other
extreme situation is when selection strength $\phi$ is $1$, which
leads to $G(\phi )=0$ and $C=0$. In this case $\qq$ goes to $1/\ab $
exponentially with time constant $-\ln (1-\ab \MutRate /(\ab
-1))\approx \ab \MutRate /(\ab -1)$.

\section{Arbitrary chromosome length $L$}
\label{APP:ExactDynamics}
In the discussion of the evolutionary
dynamics and equilibrium state, we make use of a Gaussian  (and a
Poisson distribution ) to approximate a binomial distribution,
which are valid when $L\gg 1$. In fact, one can relax this
condition and work directly with binomial distribution, with the
help of special function $I_{x}(a,b)$, the incomplete Beta
function. $I_{x}(a,b)$ is defined as
\begin{equation}
I_{x}(a,b)=\frac{1}{B(a,b)}\int _{0}^{x}dtt^{a-1}(1-t)^{b-1},
\end{equation}
where $B(a,b)$ is the complete beta function.

Suppose the probability density for a binomial distribution is
\begin{equation}
P_{\qq}(\Pheno,L)\equiv C_{L}^{\Pheno}\qq^{\Pheno}(1-\qq)^{L-\Pheno}.
\end{equation}
 The cumulative probability density is then
\begin{equation}
\sum _{i=\Pheno}^{L}P_{\qq}(i,L)=I_{\qq}(\Pheno,L+1-\Pheno)
\end{equation}

We have, at the selection,
\begin{equation}
\phi =\alpha P_{\qq}(\threshold,L)+\sum
_{i=\threshold+1}^{L}P_{\qq}(i,L), \label{EQ:BinomialConstraint}
\end{equation}
where again $\threshold$ is the threshold and the state with
$\threshold$ matches maybe partially selected.

The new individual match probability after selection and
recombination is
\begin{equation}
L\qq^{\textrm{r}}=\frac{1}{\phi }\left(\alpha \threshold
P_{\qq}(\threshold,L)+\sum
_{i=\threshold+1}^{L}iP_{\Pheno}(i,L)\right)
\end{equation}

Incorporating the mutational process and making use of
Eq.~(\ref{EQ:BinomialConstraint}), we arrive at
\begin{equation}
\qq_{t+1}=\frac{\MutRate }{\ab -1}+\frac{1-\frac{\ab }{\ab
-1}\MutRate }{L\phi }\left(\threshold\phi
+L\qq_{t}I_{\qq_{t}}(\threshold,L-\threshold)-\threshold
I_{\qq_{t}}(k\threshold+1,L-\threshold)\right),
\label{EQ:FullDescript}
\end{equation}
where we made use of the following identity
\begin{equation}
\sum_{i=\threshold+1}^{L}iP_{\qq}(i,L)=L\qq I_{\qq}(\threshold,L-\threshold).
\end{equation}

\begin{figure}
\vspace{.2cm}
\includegraphics*[  height=2in]{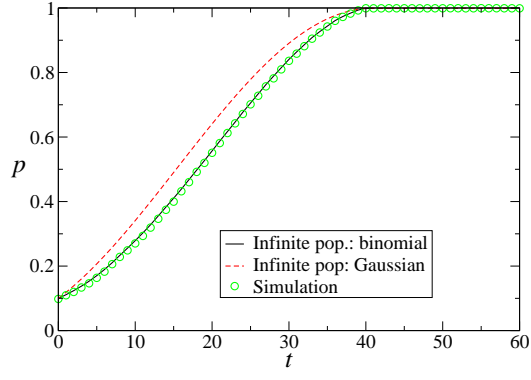}
\caption{\protect Comparison of two theoretical approaches with
simulation result. Solid curve is the result of Eq.~(\ref{EQ:FullDescript}).
Dashed curve is the result of Eq.~(\ref{EQ:Evol1}).
Here $L=10$, $\phi=.9$ and $\MutRate=.001$. In
simulation $N=10^4$. }
\label{FIG:MaxiRecombArbitraryLength}
\end{figure}

The dynamics and equilibrium properties follow from
Eq.~(\ref{EQ:FullDescript}).
Fig.~(\ref{FIG:MaxiRecombArbitraryLength}) shows that, when
chromosome length $L$ is not long, indeed this approach agrees
much better with simulation than Gaussian approximation.

\newpage

\end{document}